\documentstyle[prl,aps,twocolumn,epsf]{revtex}
\newcommand{\lapx}{\mbox{\raisebox{-4pt}{$\,\buildrel<\over\sim\,$}}} 
\newcommand{\gapx}{\mbox{\raisebox{-4pt}{$\,\buildrel>\over\sim\,$}}} 
\begin{document} 
\draft 
\title{Coulomb drag shot noise in coupled Luttinger liquids} 
\author{B.~Trauzettel$^1$, R.~Egger$^2$, and H.~Grabert$^1$} 
\address{${}^1$Fakult\"at f\"ur Physik, Albert-Ludwigs-Universit\"at, 
D-79104 Freiburg, Germany\\ 
${}^2$Institut f\"ur Theoretische Physik, 
Heinrich-Heine-Universit\"at, D-40225 D\"usseldorf, Germany} 
\date{Date: \today} 
\maketitle 
 
\begin{abstract} 
Coulomb drag shot noise has been studied theoretically for 1D interacting
electron systems, which are realized e.g.~in single-wall nanotubes. We show
that under adiabatic coupling to external leads, the Coulomb drag shot noise
of two coupled or crossed nanotubes contains surprising effects, in particular
a complete locking of the shot noise in the tubes. In contrast to Coulomb drag
of the average current, the noise locking is based on a symmetry of the
underlying Hamiltonian and is not limited to asymptotically small energy scales.
\end{abstract} 
 

\narrowtext 

Shot noise in mesoscopic systems has started to attract much interest 
\cite{blant00} because it provides information not contained in the 
current-voltage characteristics alone, and can for instance reveal the 
fractional charge of quasiparticles in exotic electronic phases such as the 
fractional quantum Hall (FQH) liquid. Shot noise in FQH bars can be
described by the chiral Luttinger liquid (LL) theory \cite{fqh}, and the 
predicted fractional charge $e^*$ smaller than the electron charge $e$ has 
been seen experimentally for filling factors $\nu=1/3$ \cite{depic97} and
$\nu=2/5$ \cite{rezn}. While in Hall bars right and left moving currents are
spatially separated, we address here current noise in {\sl non-chiral}
Luttinger liquids \cite{gogol98}, where right and left movers interact in the
same channel. Experimental realizations of this generic model for interacting
electrons in 1D are for instance provided by  single-wall
nanotubes (SWNTs)  \cite{dekker99,tube} or semiconductor quantum wires
\cite{ausl}. For clarity, we focus on SWNTs, with very similar effects
expected for other non-chiral 1D interacting metals, and restrict ourselves to
the case of spinless electrons \cite{foot}.  According to available
experiments \cite{tube},  the standard LL interaction parameter in SWNTs is
$g\approx 0.2$, indicating the presence of strong Coulomb interactions
\cite{tube2}.

The two-terminal shot noise of a nanotube with an impurity is due to
backscattered quasiparticles constructed out of the zero-modes and plasmons of
the bosonized LL theory \cite{pham} carrying a fractional charge $e^*=g
e$. Naively, one might expect that what happens at the impurity should be
observable in the shot noise, namely that the two-terminal shot noise in the
weak impurity limit can be written as $e^*$ times the backscattered current \cite{blant00}. On the other hand, as noted previously for the average
current \cite{maslov,egger96,egger00}, 
the coupling to Fermi liquid reservoirs needs
to be incorporated explicitly in the model. We demonstrate that d.c.~shot
noise does {\sl not} probe the fractional charge but coincides with the noise
in a noninteracting wire. To experimentally access the expected quasiparticle
charge then requires more complicated four-terminal setups \cite{bena00},
which seem however difficult to implement in practice and also give only indirect evidence for $e^*$.

On the other hand, pronounced interaction effects in shot noise experiments
arise from the Coulomb drag shot noise of two adjacent
or crossed SWNTs, where we predict the remarkable effect of {\sl complete noise
locking}. In fact, the noise power $P_1$ in one SWNT induces a
Coulomb drag shot noise $P_2$ in the other SWNT (and vice versa), where
both are perfectly locked together, $P_1=P_2$. This phenomenon would not show
up in Fermi liquid systems \cite{gurevich} and is only based on an emergent
symmetry of the interacting low-energy theory. As a consequence, it can be
observed at arbitrary temperature $T$ and applied voltage $U$ (within the
range of validity of the LL model). Hence, the predicted complete noise
locking is a basic manifestation of interaction effects with a more
fundamental origin than the absolute drag effect of the average current
\cite{flensberg,nazaro,komnik}. In contrast to the noise, the absolute drag of
the current is not based on a symmetry of the Hamiltonian and is therefore
restricted to asymptotically small energy scales $k_B T, eU$ and a large
contact region between the tubes. Recently, current drag effects have been
observed for crossed SWNTs \cite{crossed} and parallel semiconductor quantum
wires \cite{debray}. Essentially the same systems should also reveal the
Coulomb drag shot noise predicted here. In future nanoscale electronic devices
with 1D transmission lines the extreme noise sensitivity of these lines should
play an important role in their applicability. Thus, a proper understanding of
their noise properties will be essential for a successful development of
future electronic components.

The problem of Coulomb drag shot noise in two coupled SWNTs maps onto two
problems of two-terminal shot noise in the 
presence of an impurity. Therefore, we
first address the physics of that scenario before we return to the
consequences on Coulomb drag shot noise after the two-terminal noise result in
Eq.~(\ref{sn3}) below. We consider a SWNT of length
$L$ that contains an impurity at $x=0$ and is adiabatically coupled to external leads. Using integrability techniques the
current $I$ through such a systems has recently been calculated exactly
\cite{egger00}. The result is $I=(e^2/h)(U-V)$, where the four-terminal
voltage $V$ has to be determined self-consistently for given $g$, impurity
strength $\lambda$, and two-terminal voltage $U$. The {\sl two-terminal current noise}, commonly measured in the leads, $|x|>L/2$, is defined  by 
\begin{equation} \label{sn1} 
P  = \int_{-\infty}^\infty dt \ e^{i\omega t} \left\langle 
\left\{ \Delta I(x,t) , \Delta I(x,0) \right\}_+ \right\rangle \;, 
\end{equation} 
where $\Delta I(x,t) = I(x,t)-\langle I \rangle$ is the current fluctuation 
operator. Since our calculations for $k_B T \gg eU$ reproduce the expected 
Johnson-Nyquist noise \cite{blant00}, let us focus for the moment on the zero
temperature limit. As is elaborated below, in the weak impurity limit the d.c.~shot noise is given by $P=2 e I_{\rm BS}$, with the
backscattered current $I_{\rm BS}=(e^2/h)U-I=(e^2/h)V$ and, similarly, in the
opposite strong backscattering limit, $P= 2 e I$. This means that it is
impossible  to observe the fractional quasiparticle charge $e^*=ge$ in a
two-terminal measurement. In a sense, the low-frequency shot noise is
completely determined by the charge $e$ of a lead electron scattered by the
compound LL-plus-impurity. We remark that for finite frequency $\omega$, the
current noise contains more information, including spatially dependent
oscillatory behaviors. Unfortunately, however, the condition $\omega L/ v_F
\gapx 1$ with the Fermi velocity $v_F$, typically requires to study
frequencies above $\approx 10 \ {\mbox {GHz}}$ in order to get significant
deviations from the low frequency result. As such high frequencies are difficult to reach experimentally, we focus on the d.c.~limit. 
 
A computation of the shot noise for this problem faces conceptual
difficulties, since, as mentioned above, the proper coupling to the Fermi
liquid leads is esssential. To obtain the current, it is sufficient to specify the mean
densities of right- or left-moving electrons injected from the reservoirs,
leading to radiative  boundary conditions \cite{egger96} that keep
integrability intact \cite{egger00}.  To access the shot noise, however, one
also needs to  simultaneously specify fluctuation properties in the leads.
A simple approach to accomplish this is to employ the inhomogeneous LL model
\cite{maslov} characterized by an $x$-dependent LL interaction parameter
$g(x)$, with $g(|x|<L/2)=g$ in the SWNT and $g=1$ in the leads. Such
modelling, when combined with a careful analysis of the electrostatics under
an applied voltage, is able to reproduce the Johnson-Nyquist equilibrium noise 
which a naive direct LL calculation does violate. Unfortunately, exact results
for the shot noise at arbitrary impurity strength seem elusive for such a
complicated model,  but to lowest order in the impurity strength, the
analytical solution is possible and outlined next.

The Hamiltonian is $H=H_0+H_U+H_I$, where $H_U$ describes the coupling to
reservoirs, $H_I$ the impurity, and the inhomogeneous LL Hamiltonian is  ($\hbar=1$) 
\begin{equation}\label{h0} 
H_0 = \frac{v_F}{2} \int dx\, \left[ \Pi^2 + \frac{1}{g^{2}(x)} (\partial_x \theta )^2 \right] \; , 
\end{equation} 
where $\Pi(x)$ is the canonical momentum to the standard boson field
$\theta(x)$ under bosonization \cite{gogol98}. Along the lines of
Ref.~\cite{maslov}, we obtain the boson correlator
$C(x,t)=\langle\theta(x,t)\theta(0,0)\rangle$ of the clean system in closed form, 
\begin{eqnarray} \label{cxt} 
&& C(x,t) =  
\int_0^\infty \frac{d\omega}{4\pi\omega}  
\sum_{s=\pm} \frac{gs e^{i\omega t}}{\kappa_s - W \kappa_{-s}} \\ 
&\times& \left\{ \begin{array}{r@{\quad \;, \quad}l} 2 e^{i s \omega|x|/v_F} 
W^{(1-s/g)/2} & |x|>L/2 \\  
\kappa_s e^{i\omega g|x|/v_F} + \kappa_{-s} W e^{-i\omega g|x|/v_F} & |x|<L/2 
\end{array} \right. \nonumber
\end{eqnarray} 
where $\kappa_\pm = 1\pm g$ and $W=\exp(ig \omega L/v_F)$. Impurity
backscattering at $x=0$ then gives $H_I = \lambda \cos [ \sqrt{4 \pi} \theta
(0) ]$ with the impurity strength $\lambda$, while impurity forward scattering
only leads to an additional phase factor of the electron field operator which
does not affect the current noise and is omitted henceforth. Since a $g=1$ LL
is a Fermi gas and not a Fermi liquid, to describe the coupling to the
applied voltage $U$, it is essential to take into account electroneutrality in
the leads. Electroneutrality implies a shift of the band bottoms strictly
following the chemical  potentials. Effectively, the electrostatic potentials
are then $eU(x<-L/2)=\mu_L$ and $eU(x>L/2)=\mu_R$ for chemical potentials
$\mu_{L/R}$ in the left ($L$) and right ($R$) reservoir, where we put
$\mu_L=eU/2$  and $\mu_R=-eU/2$. In contrast to the leads, the SWNT does not
remain electroneutral in presence of an applied voltage
\cite{egger96}. This is a consequence of the finite range of the internal
Coulomb interaction screened by external gates. The band bottom in the SWNT is
also shifted with a jump of size $eV$ at the impurity site.  As shown in
Ref.~\cite{egger96}, the electrostatics within a voltage biased SWNT emerges
naturally as a steady-state interaction effect with $U(x)=0$ for
$|x|<L/2$. The applied voltage $U$ then gives rise to the contribution 
\begin{equation}\label{hu} 
H_U = \frac{e}{\sqrt{\pi}} \int dx \ U(x) \partial_x \theta \; . 
\end{equation} 
This formulation does not make the (incorrect) assumption of a local drop of
the applied voltage $U$ at the impurity site. In reality, this potential drop
is given by the four-terminal voltage $V$, and it is essential to distinguish $V$ and $U$.
 
To compute the shot noise to lowest order in $\lambda$, we then proceed along
the lines of the Keldysh formalism \cite{rammer}. Expanding in $\lambda$ and
using the boson correlator $C(x,t)$ in Eq.~(\ref{cxt}) gives after
straighforward yet tedious algebra for the shot noise measured in the leads
\begin{equation} \label{sn3} 
P = 2 e^2 \lambda^2 \sin(\pi g) \cos(\pi g) \Gamma(1-2g) 
\frac{(eU)^{2g-1}}{\omega_c^{2g}} \;,  
\end{equation} 
up to corrections vanishing for $L \rightarrow \infty$. The same finite $L$
corrections also appear in the backscattered current $I_{\rm BS}$ of the
inhomogenous LL system and therefore do not affect the final current-noise
relation, $P = 2 e  I_{\rm BS}$. This result for weak impurity backscattering
shows that it is indeed the electron charge $e$ and not the fractional charge
$e^*=ge$ that governs the d.c.~shot noise spectrum in a SWNT adiabatically
coupled to external leads. Although  the relation $P=2e I_{\rm BS}$ has been
proposed previously \cite{ponom99}, the derivation has been questioned, see,
e.g., Ref.~\cite{blant00}. We believe that our careful treatment resolves this
issue. In the opposite limit of a strong impurity, the contribution $H_I$ is
replaced by a term $H_T$ describing electron tunneling through the ``barrier''
region. Then the expected  relation $P=2eI$ can be easily derived
perturbatively in $H_T$, in analogy to our treatment of the weak-impurity limit.

\begin{figure} 
\hspace{0.5cm}
\epsfysize=4cm 
\epsffile{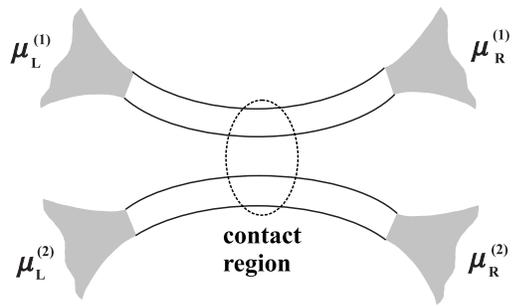} 
\vspace{0.3cm} 
\caption{Two coupled SWNTs with a local 
contact region of length $L_c$.  The SWNTs are adiabatically
connected to separate reservoirs.} 
\label{twoqws} 
\end{figure}

Now we turn to the more interesting problem of {\sl Coulomb drag shot noise},
which has immediate application potential for SWNTs. The setup under consideration
involves two SWNTs that are separately contacted with applied voltages
$U_{1,2}$, but remain in contact over some region of 
length $L_c$.  Again we assume good (adiabatic) contact with the leads,
which is possible using present-day technology \cite{dekker99}.
The contact between the two SWNTs
may be achieved either by arranging the SWNTs parallel
to each other as illustrated schematically in Fig.~\ref{twoqws},
implying $L_c\gg a$ with the lattice spacing $a$, or by
crossing them with $L_c\approx a$. 
We note that crossed SWNTs in good contact to separate lead
electrodes have already been realized
experimentally \cite{crossed},
and a measurement of noise properties can be anticipated
without major difficulties.
The question then arises: What is the shot noise
transferred via the Coulomb interaction coupling 
the two SWNTs (``Coulomb drag shot noise'')?   
As we discuss below, the combined presence 
of intra- and inter-tube interactions gives rise 
to interesting shot noise phenomena. 
We focus on the case of
short-to-intermediate contact length $L_c$.  Therefore our treatment
directly applies to crossed SWNTs, but also gives a correct 
qualitative picture for parallel SWNTs.  This can be
understood from a renormalization group calculation \cite{komnik},
showing that the low-energy properties of an extended contact  are captured
by an effective pointlike contact (with renormalized coupling
strength) for not exceedingly large $L_c$.  
In practice, a contact with $L_c \lapx 20 a$
can still  be modelled by such  a pointlike
contact to high accuracy. 
 
Taking adiabatic contacts to the leads, the 
two uncoupled clean SWNTs ($\alpha=1,2$) are described by the  
above Hamiltonian, $H_\alpha= H_0+H_{U_\alpha}$, where 
we take the same $g$ parameter for both SWNTs. 
Since the band curvature is extremely small \cite{dekker99},
both Fermi velocities are identical, $v_{F1}=v_{F2}=v_F$, and
independent of the Fermi momentum $k_F=|E_F|/v_F$.  The Fermi level can in practice be 
tuned separately for both SWNTs by changing the mean chemical
potential in the attached leads, and therefore in general $k_{F1}\neq k_{F2}$.
We assume that the inter-tube coupling 
is sufficiently weak  such that neither
electron backscattering within each SWNT alone nor
electron tunneling between the SWNTs is significant. 
In any case, the strong interactions present in SWNTs ($g\approx 0.2$)
highly suppress tunneling into or between them, and recent experiments
\cite{crossed} are indeed consistent with very small single-electron tunneling
probabilities.  
However, it is crucial to retain the {\sl electrostatic coupling}
between the SWNTs. 
Modelling this coupling by a local interaction in the contact region
$-L_c/2<x<L_c/2$  as indicated in Fig.~\ref{twoqws},
we obtain the contribution  \cite{foot2}
\begin{eqnarray}\label{hiw} 
H' &=& \int dx \zeta(x) \Bigl\{ V_{\rm FS} \prod_\alpha \partial_x
\theta_\alpha \nonumber \\ 
&& + V_{\rm BS} \prod_\alpha \sin[ 2 k_{F\alpha} x +\sqrt{4\pi} 
\theta_\alpha (x)] \Bigr\} \; , 
\end{eqnarray}
where $\zeta(x)=1$ in the contact region and zero otherwise, 
and $V_{\rm FS}$ ($V_{\rm BS}$) is the forward (backward) scattering
inter-tube interaction strength.  
The $V_{\rm FS}$ term has no effect on noise nor current
for a local coupling (small $L_c$), and gives only an insignificant 
renormalization of LL parameters  for large $L_c$
that will be ignored here.

Progress can then be made by switching to new boson fields 
$\theta_\pm = (\theta_1 \pm \theta_2)/\sqrt{2}$, which completely 
decouple the full Hamiltonian $H=H_1+H_2+H'=H_++H_-$ in the 
$\pm$ channels,
\begin{eqnarray} \label{hpm}
H_\pm &=& \frac{v_F}{2} \int dx \ \left[ \Pi_\pm^2 + \frac{1}{g^{2}(x)}
(\partial_x \theta_\pm)^2 \right] \\
&+& \frac{e}{\sqrt{\pi}} \int dx \  U_\pm (x) \partial_x \theta_\pm \nonumber
\\ 
&\mp& \frac{V_{\rm BS}}{2} \int dx \zeta(x) \cos [ 2(k_{F1}\pm
k_{F2})x+\sqrt{8\pi} \theta_\pm (x) ] \nonumber \;.
\end{eqnarray}
Now we make use of the small-$L_c$ assumption which
allows to formulate the last line of Eq.~(\ref{hpm}) as an effectively
point-like coupling $\mp \lambda_\pm \cos[\sqrt{8\pi}\theta_\pm(0)]$. The
effective impurity strength in the two channels is then
\begin{equation}\label{jpm} 
\lambda_\pm = \frac{V_{\rm BS}}{2} \int dx \zeta(x) \cos [ 2 ( k_{F1} \pm k_{F2}) x] \; . 
\end{equation} 
Notably, these couplings could be tuned by either (i) changing the distance
between the two SWNTs which modifies $V_{\rm BS}$, (ii) changing the contact
length $L_c$, or (iii) changing the Fermi momenta $k_{F, 1/2}$ relative 
to each other. Such a {\sl tunable impurity strength} gives an exceptional
 flexibility in transport measurements. Furthermore, the electrostatic
potentials $U_\pm(x)$ are of the same form as before, but with effective
applied voltages  $U_\pm = (U_1 \pm U_2)/\sqrt{2}$. We are then left with
two independent effective single-impurity problems in the $\pm$ channels. Due
to the doubled argument in front of $\theta_\pm(0)$ in the impurity cosine, 
these models live at the doubled interaction parameter, $g\to 2 g$. 

Applying this transformation shows that the shot noise $P_\alpha$ in each physical SWNT ($\alpha=1,2$) is {\sl identical}. Formally, this
follows from the vanishing of cross-correlations between the $\theta_+$ and
$\theta_-$ fields. In terms of the shot noise $P_\pm$ in the $\pm$ channel, 
which for small $V_{\rm BS}$ is given by Eq.~(\ref{sn3}) with $g\to 2g$,
$\lambda\to \lambda_\pm$, $U\to U_\pm$,
\begin{equation} \label{na} 
P_1 = P_2 =  ( P_+  + P_- )/2  \; . 
\end{equation} 
The predicted {\sl complete locking of shot noise}, $P_1=P_2$, is solely
based on the decoupling $H=H_+ + H_-$, and therefore survives
thermal fluctuations and finite applied voltages. As the decoupling,
Eq.~(\ref{hpm}), is valid for any contact length $L_c$, the noise locking is
also observable in parallel SWNTs with an extended contact region. Of
course, the noise locking is affected by inter-tube tunneling processes, but
as these are highly suppressed for strong intra-tube interactions (small $g$),
Eq.~(\ref{na}) should be valid to high accuracy. Note that noise locking
happens already for a very short contact, $L_c \approx a$, while the
transconductance and hence the absolute drag effect would vanish in that limit \cite{flensberg}.

To demonstrate the consequences of changing $\lambda_\pm$ relative to each
other, we finally point out an interesting feature for $\lambda_+\ll
\lambda_-$. This situation applies to the case of parallel SWNTs characterized
by intermediate-to-long $L_c$, where $\lambda_+$ is suppressed by the
oscillating integrand in Eq.~(\ref{jpm}) and can be treated perturbatively,
while $\lambda_-$ is quite large and can even reach the strong-impurity 
limit.  Exploiting duality relations \cite{fqh}, the shot noise $P=P_1=P_2$ is
found from Eq.~(\ref{na}) as $P = e( I_+^{\rm BS} + I_- )$, 
where $I_+^{\rm BS} \sim \lambda_+^2 |U_1+U_2|^{4g-1}$ and $I_- \sim
\lambda_-^{-1/g}|U_1-U_2|^{1/g-1}$. Apparently, for $U_1=U_2$, shot noise is
only determined by the $+$ channel and does not depend on $\lambda_-$ at all.
For $U_1 \neq U_2$,  however,  one can reach the strong-impurity limit in the
$-$ channel. Using Eq.~(\ref{jpm}) we then predict a $g$-dependent power-law
decrease of $P$ as a function of the inter-tube coupling, $P\sim V_{\rm
BS}^{-1/g}$.  As a consequence, there should be a {\sl maximum} in the
noise as a function of $V_{\rm BS}$ at some (unknown) intermediate
value.  
 
In summary, we have shown that the Coulomb drag shot noise of two coupled
SWNTs reveals unambiguous evidences for inter- and intra-tube interaction
effects. In particular, we have predicted a complete noise locking in each
physical tube due to an emergent symmetry of the Hamiltonian. The system under
consideration may be realized experimentally by crossing two SWNTs with a remarkable tunability of the effective inter-tube coupling strength.

We thank A.~De Martino, A.~Komnik and H.~Saleur for discussions.
 This work has been supported by the DFG under Grant No. GR 638/19-1.

\begin{references}
\bibitem{foot} 
According to the low-energy theory \cite{tube2},
spinless results are easily adapted to include the spin
and K point degeneracy. Hence our conclusions apply generally.
\bibitem{foot2}
Up to weak (multiplicative) logarithmic corrections,
our results also hold for $1/r$ interactions.
\end{references}
\end{document}